\documentclass[11pt,a4paper]{PoS}

\newcommand{\nc}{\newcommand}
\nc{\am}[1]{\ensuremath{#1}}

\nc{\ke}{\am{K \to e \nu}}
\nc{\km}{\am{K \to \mu \nu}}
\nc{\kpd}{\am{K_{\pi 2}}}
\nc{\ked}{\am{K_{e2}}}
\nc{\kedg}{\am{K_{e2\gamma}}}
\nc{\ket}{\am{K_{e3}}}
\nc{\kmt}{\am{K_{\mu3}}}

\nc{\kmd}{\am{K_{\mu2}}}
\nc{\erreg}{\am{R_{\gamma}}}
\nc{\kee}{\am{K \to \pi  e \nu}}
\nc{\kmm}{\am{K \to \pi \mu \nu}}
\nc{\nn}{\am{N\!N}}
\nc{\NN}{\am{N\!N}}

\nc{\be}{\begin{equation}}
\nc{\ee}{\end{equation}}
\nc{\ChPT}{$\chi$PT}
\nc{\DAF}{DA\char8NE}
\nc{\GeV}{\mbox{GeV}}
\nc{\ps}{\mbox{ps}}
\nc{\mrad}{\mbox{mrad}}
\nc{\ie}{i.e.}
\nc{\MeV}{\mbox{MeV}}
\nc{\mbo}{\mathversion{bold}}

 \def\f{$\phi$}  \def\ab{$\sim$}  
    \def\gam{\gamma}
\def\vp{{\vphantom{$I^{I^i}$}}}

\renewcommand\Gamma{\char0}
\renewcommand\Delta{\char1}

\title{Test of lepton flavor violation with Ke2 decay at KLOE }

\ShortTitle{LFV test with Ke2 decay at KLOE}

\author{\speaker{Barbara Sciascia}\thanks{for the KLOE Collaboration:
F.~Ambrosino, A.~Antonelli, M.~Antonelli, F.~Archilli, P.~Beltrame, 
G.~Bencivenni, C.~Bini, C.~Bloise, S.~Bocchetta, F.~Bossi, P.~Branchini, 
G.~Capon, D.~Capriotti, T.~Capussela, F.~Ceradini, P.~Ciambrone, E.~De Lucia, 
A.~De Santis, P.~De Simone, G.~De Zorzi, A.~Denig, A.~Di Domenico, C.~Di Donato, 
B.~Di Micco, M.~Dreucci, G.~Felici, S.~Fiore, P.~Franzini, C.~Gatti, P.~Gauzzi, 
S.~Giovannella, E.~Graziani, M.~Jacewicz, V.~Kulikov, G.~Lanfranchi, J.~Lee-Franzini, 
M.~Martini, P.~Massarotti, S.~Meola, S.~Miscetti, M.~Moulson, S.~M\"uller, F.~Murtas, 
M.~Napolitano, F.~Nguyen, M.~Palutan, A.~Passeri, V.~Patera, P.~Santangelo, B.~Sciascia, 
A.~Sibidanov, T.~Spadaro, M.~Testa, L.~Tortora, P.~Valente, G.~Venanzoni, and R.~Versaci.}\\
        LNF-INFN\\
        E-mail: \email{barbara.sciascia@lnf.infn.it}}

\abstract{  We present a precise measurement of the ratio
  $R_K=\Gamma(K\to e\nu(\gamma))/\Gamma(K\to \mu\nu(\gamma))$ performed with the
  KLOE detector.  The results are based on data
  collected at the Frascati $e^+e^-$ collider \DAF\ for an integrated
  luminosity of 2.2 fb$^{-1}$.
  We find $R_K = (2.493\pm0.025_\mathrm{stat}\pm0.019_\mathrm{syst})\times 10^{-5}$,
  in agreement with the Standard Model expectation. This result is
  used to improve constraints on parameters of the Minimal
  Supersymmetric Standard Model with lepton flavor violation.}

\FullConference{2009 KAON International Conference KAON09,\\
		 June 09 - 12 2009\\
		 Tsukuba, Japan}

\begin{document}
The decay $K^\pm\!\to e^\pm\nu$ is strongly suppressed,
$\sim$few$\times$10$^{-5}$, because of conservation of angular
momentum and the vector structure of the charged weak current. It
therefore offers the possibility of detecting minute contributions from
physics beyond the Standard Model (SM). This is particularly true of the
ratio $R_K=\Gamma(\ke)/\Gamma(\km)$ which, in the SM, is
calculable without hadronic uncertainties \cite{marci,Cirigliano:2007xi}.
Physics beyond the SM, for example mul\-ti\--Higgs effects inducing an effective
pseudo-scalar interaction, can change the value of $R_K$.
It has been shown in Ref. \cite{masiero} that deviations of $R_K$ of
up to {\em a few percent} are possible in minimal
supersymmetric extensions of the SM (MSSM) with non vanishing $e$-$\tau$
scalar lepton mixing. To obtain
accurate predictions, the radiative process $K\to e \nu \gamma$ (\kedg)
must be included. In \kedg, photons can be produced via
internal-bremsstrahlung (IB) or  direct-emission (DE), the latter being
dependent on the hadronic structure.  Interference among the two
processes  is negligible \cite{bijnens}.  The DE contribution to the
total width is approximately equal to that of IB \cite{bijnens} but is
presently known with a 15\% fractional accuracy \cite{Heintze:1977kk}.

$R_K$ is {\em defined} to be inclusive of IB, ignoring however
DE contributions.  A recent calculation \cite{Cirigliano:2007xi},
which includes order $e^2p^4$ corrections in chiral perturbation
theory (\ChPT), gives:
\begin{equation}
\label{eq:rksm}
 R_K = (2.477\pm0.001)\times 10^{-5}.
\end{equation}
$R_K$ is not directly measurable, since
IB cannot be distinguished from DE on an event-by-event basis.
Therefore, in order to compare data with the SM prediction at the
percent level or better, one has to be careful with the DE part.
We define the rate $R_{10}$ as:
\be
R_{10}=\Gamma(\ke(\gamma),\ E_\gamma<10\mathrm{\ MeV})/\Gamma(\km).
\label{eq:R10}
\ee
Evaluating the IB spectrum to ${\mathcal O}(\alpha_{\rm em})$
with resummation of leading logarithms, $R_{10}$ includes
$93.57\pm0.07\%$ of the IB,
\be
R_{10}=R_K\times(0.9357\pm0.0007).
\label{eq:R10new}
\ee
The DE contribution in this range is expected to be negligible.
$R_{10}$ is measured without photon detection. 
Some small contribution of DE is present in the selected sample.
In particular, DE decays have some overlap
with the IB emission at high $p_e$.
We have also measured \cite{kaon09:moulson} the differential width
$\mathrm{d}R_\gamma/\mathrm{d}E_\gamma$ for $E_\gamma\!>10$ MeV and $p_e\!>200$ MeV
requiring photon detection, both to test \ChPT\ predictions for the
DE terms and to reduce possible systematic uncertainties on the
$R_{10}$ measurement.

\section{Selection of leptonic kaon decays}
\label{sec:method}
\begin{figure*}[ht]\centering
  \includegraphics[width=0.4\linewidth]{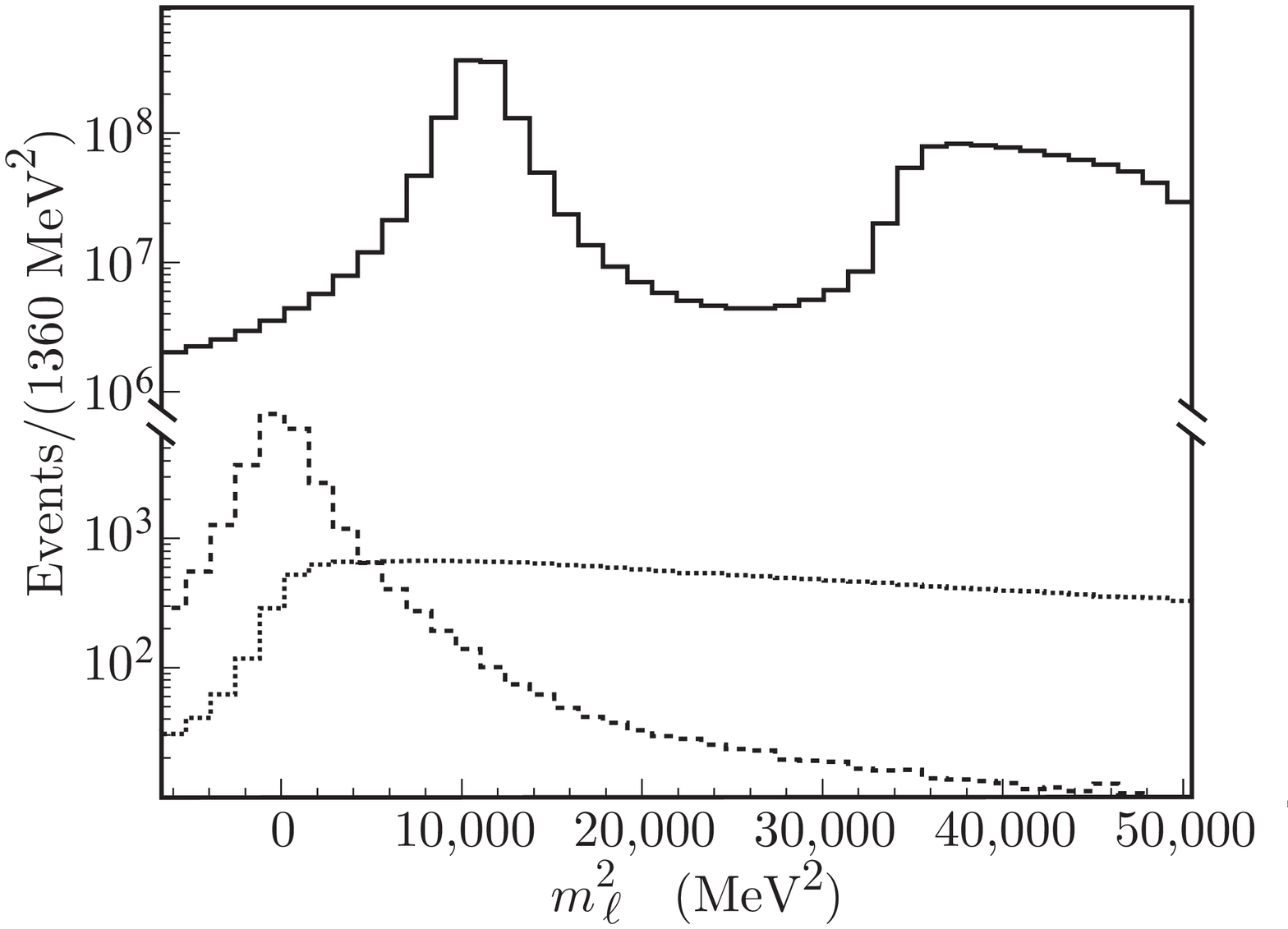}
  \includegraphics[width=0.4\linewidth]{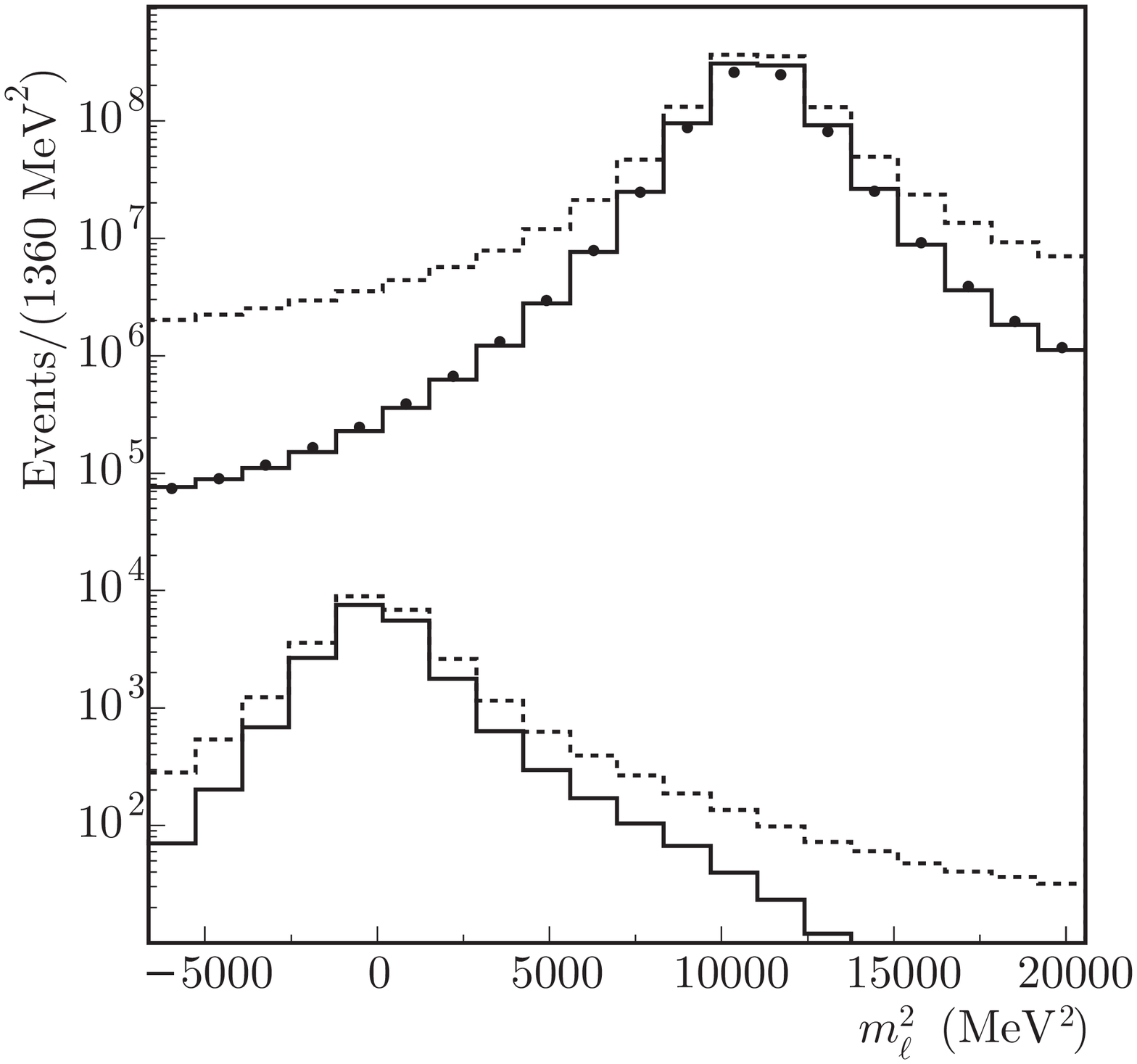}
  \caption{Left: MC distribution of $m_\ell^2$, solid
    line. The contribution of \kedg\ with $E_\gamma< 10$ MeV
    ($>10$ MeV) is shown by the dashed (dotted) lines.
    Right: $m_\ell^2$ spectrum before (dashed) and
    after (solid) quality cuts for MC \kmd\ (upper plots)and 
    \ked\ with $E_\gamma<10$ MeV (lower plots). Black dots represent data
    after quality cuts.}
  \label{fig:mlep}
\end{figure*}

$K^\pm$ decays are signaled by the observation of two tracks
with the following conditions. One track must originate at the interaction point (IP) and have
 momentum in the interval $\{70,\:130\}$ MeV, consistent with being a kaon from \f-decay. The
second track must originate at the end of the previous track and have momentum larger than that
of the kaon, with the same charge. The second track is taken as a decay product of the kaon. The
point of closest approach of the two tracks is taken as the kaon decay point D and must satisfy
40$<r_{\rm D}<$150 cm, $|z_{\rm D}|<$80 cm. The geometrical acceptance with these conditions
is \ab56\%, while the decay point reconstruction efficiency is \ab51\%. From the measured kaon
and decay particle momenta, ${\bf p}_K$ and ${\bf p}_{\rm d}$, we compute the squared mass
$m_\ell^2$ of the lepton for the decay $K\to\ell\nu$ assuming zero missing mass:
\be
m_\ell^2=\left(E_K-\left|{\bf p}_K-{\bf p}_{\rm d}\right|\right)^2-{\bf p}_{\rm d}^2{\mbox .}
\label{eq5}
\ee
The distribution of $m_\ell^2$ is shown in Fig. \ref{fig:mlep} left panel, upper curve, from MC simulation.
The muon peak is quite evident, higher masses corresponding to non leptonic and semileptonic
decays. No signal of the $K\to e\nu$ (\ked) decay is vi\-si\-ble.
The very large background around zero mass
is the tail of the \km\ (\kmd) peak, due to poor measurements of $p_K$,
$p_{\rm d}$ or the decay angle, $\alpha_{K\rm d}$.
The expected signal from \kedg\ is also shown in Fig. \ref{fig:mlep} left, lower curves, separately
for $E_\gamma>$10 and $<$10 MeV. The expected number of \ked\
decays in the sample is \ab30,000. A background rejection of at least 1000 is necessary, to
obtain a 1\% precision measurement of $\Gamma(\ked)$, with an efficiency of \ab30\%.

The kinematics of the two-body decay $\phi\to K^+K^-$ provides an
additional measurement of $p_K$
with $\sim1$ \MeV\ resolution, comparable with that from track
reconstruction. We require the two $p_K$ determinations to agree within 5 MeV.
Further cuts are applied to the daughter track. Resolution of track
parameters is improved by rejecting badly reconstructed tracks, i.e.,
with $\chi^2({\rm track\ fit)}/\mathrm{ndf}\!>\!7.5$. 
Finally, using the expected errors on $p_K$ and
$p_{\rm d}$ from tracking, we compute event by event the error on $m_\ell^2$,
$\delta m_\ell^2$. 
The distribution of $\delta m_\ell^2$ depends slightly on the opening angle $\alpha_{Kd}$,
which in turn has different distribution for \ked\ and \kmd.
Events with large value of $\delta m_\ell^2$ are rejected: $\delta m_\ell^2<\delta_\mathrm{max}$, with
$\delta_\mathrm{max}$ defined as a function of $\alpha_{Kd}$,
to equalize the losses due to this cut for \ked\ and \kmd.
The effect of quality cuts on $m_\ell^2$ resolution is shown
in Fig. \ref{fig:mlep}, right.  The background in the \ked\ signal region is
effectively reduced by more than one order of magnitude with an
efficiency of $\sim$70\% for both \ked\ and \kmd.

\begin{figure*}[ht]\centering
  \includegraphics[width=0.35\linewidth]{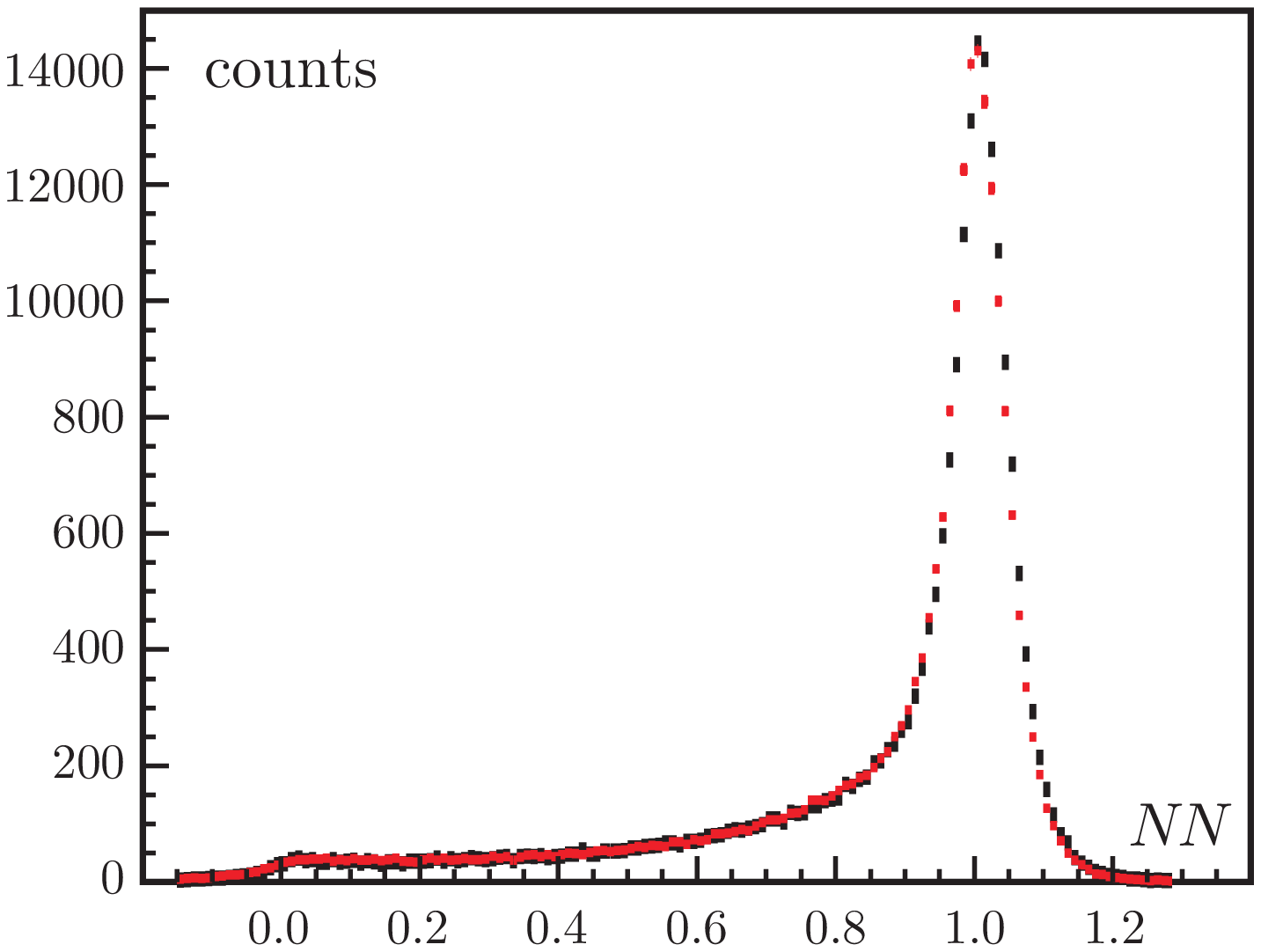}
  \includegraphics[width=0.3\linewidth]{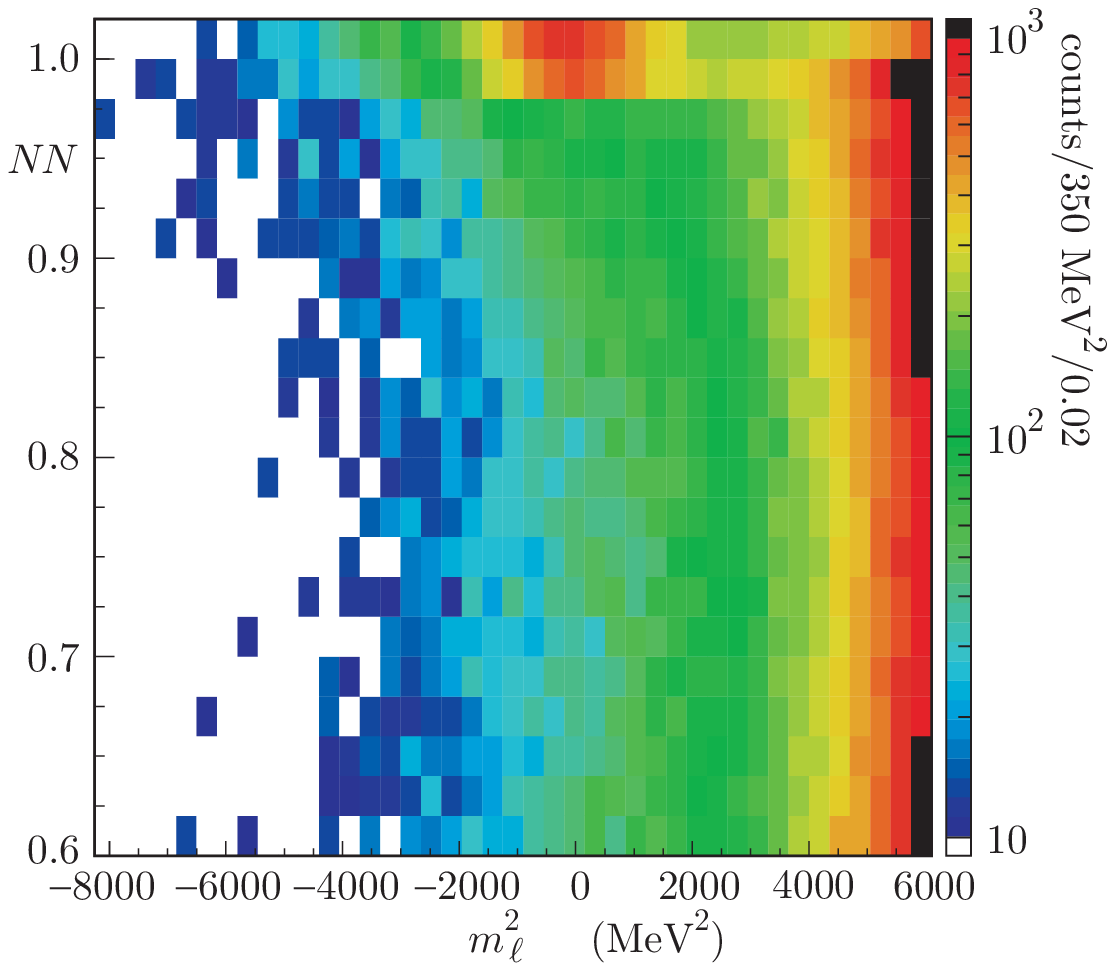}
  \caption{Left: Neural-network output, \nn, for electrons of a $K_L\to \pi e \nu$
   sample from data (black) and MC (red).
   Right: Data density in the \nn, $m_\ell^2$ plane.}
  \label{ke2:pidNN}
\end{figure*}

Information from the EMC is used to improve background rejection. To
this purpose, we extrapolate the secondary track to the EMC surface
and associate it to a nearby EMC cluster.  This requirement produces a
signal loss of about 8\%.
Energy distribution and position along the shower axis of all cells
associated to the cluster allow for $e/\mu$ particle identification.
For electrons, the cluster energy $E_\mathrm{cl}$ is a
measurement of the particle momentum $p_d$, so that
$E_\mathrm{cl}/p_d$ peaks around 1, while for muons
$E_\mathrm{cl}/p_d$ is on average smaller than 1.
Moreover, electron clusters can also be distinguished from $\mu$
(or $\pi$) clusters by exploiting the granularity of the EMC.
All useful information about shower profile and total energy deposition are
combined with a 12-25-20-1 structure neural network trained on
$K_L\to \pi \ell \nu$ and \kmd\ data, taking into
account variations of the EMC response with momentum and impact angle
on the calorimeter. The distribution of the neural network output, \nn, for a
sample of  $K_L\to \pi e \nu$ events is shown in Fig. \ref{ke2:pidNN} left,
for data and MC. Additional separation has been obtained using time
of flight information.
The data distribution of \nn\ as function of $m_\ell^2$ is
shown in Fig. \ref{ke2:pidNN} right. A clear \ke\ signal can be seen
at $m_\ell^2\sim0$ and $\nn\sim1$.

Some $32\%$ of the events with a $K$ decay in the fiducial volume,
have a reconstructed kink matching the required quality criteria {\em and} an EMC cluster
associated to the lepton track; this holds for both \ked\ and
\kmd. In the selected sample, the contamination from $K$ decays other than $K_{\ell2}$ is negligible, as
evaluated from MC.  
$R_{10}$, Eq. \ref{eq:R10}, is obtained without requiring the presence of the radiated photon. 
The number of $\ke(\gamma)$, is determined with a binned
likelihood fit to the two-dimensional \nn\ vs $m_\ell^2$ distribution.
Distribution shapes for signal and \kmd\ background are taken from MC; the normalization
factors for the two components are the only fit parameters. The fit has been performed
in the region $-3700<m_\ell^2<6100$ MeV$^2$ and $\nn>0.86$.
The fit region accepts $\sim90\%$ of $\ke(\gamma)$ events with $E_\gamma<10$ MeV, as evaluated from MC.
A small fraction of fitted $\ke(\gamma)$ events have $E_\gamma>10$ MeV: the value of this
``contamination'', $f_{\rm DE}$, is fixed in the fit to the expectation from simulation,
$f_\mathrm{DE} = 10.2\%$. A systematic error related to this assumption is discussed in Sect. \ref{sec:cs}.

\begin{figure*}[ht]\centering
    \includegraphics[width=0.35\textwidth]{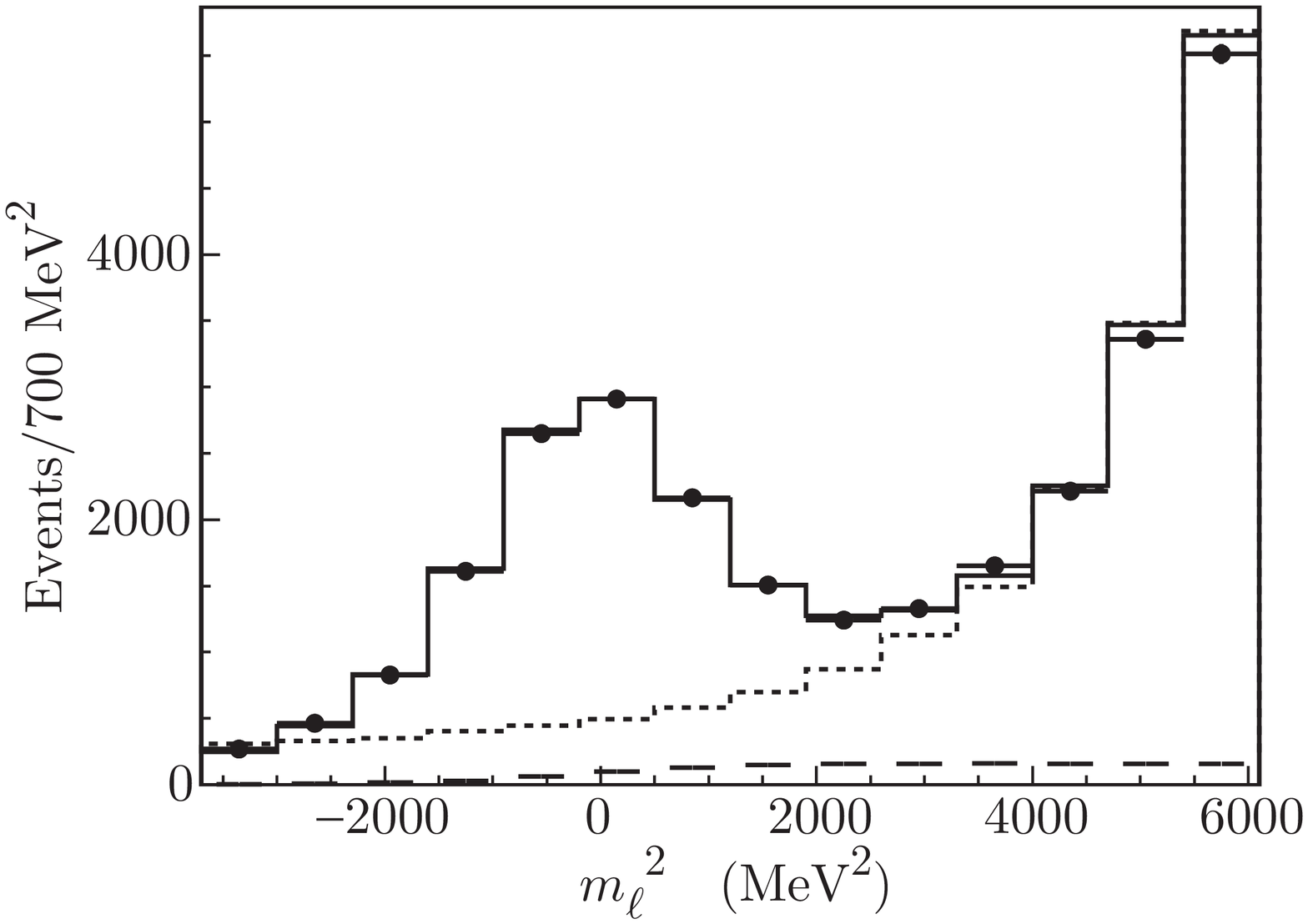}\kern1cm
    \includegraphics[width=0.35\textwidth]{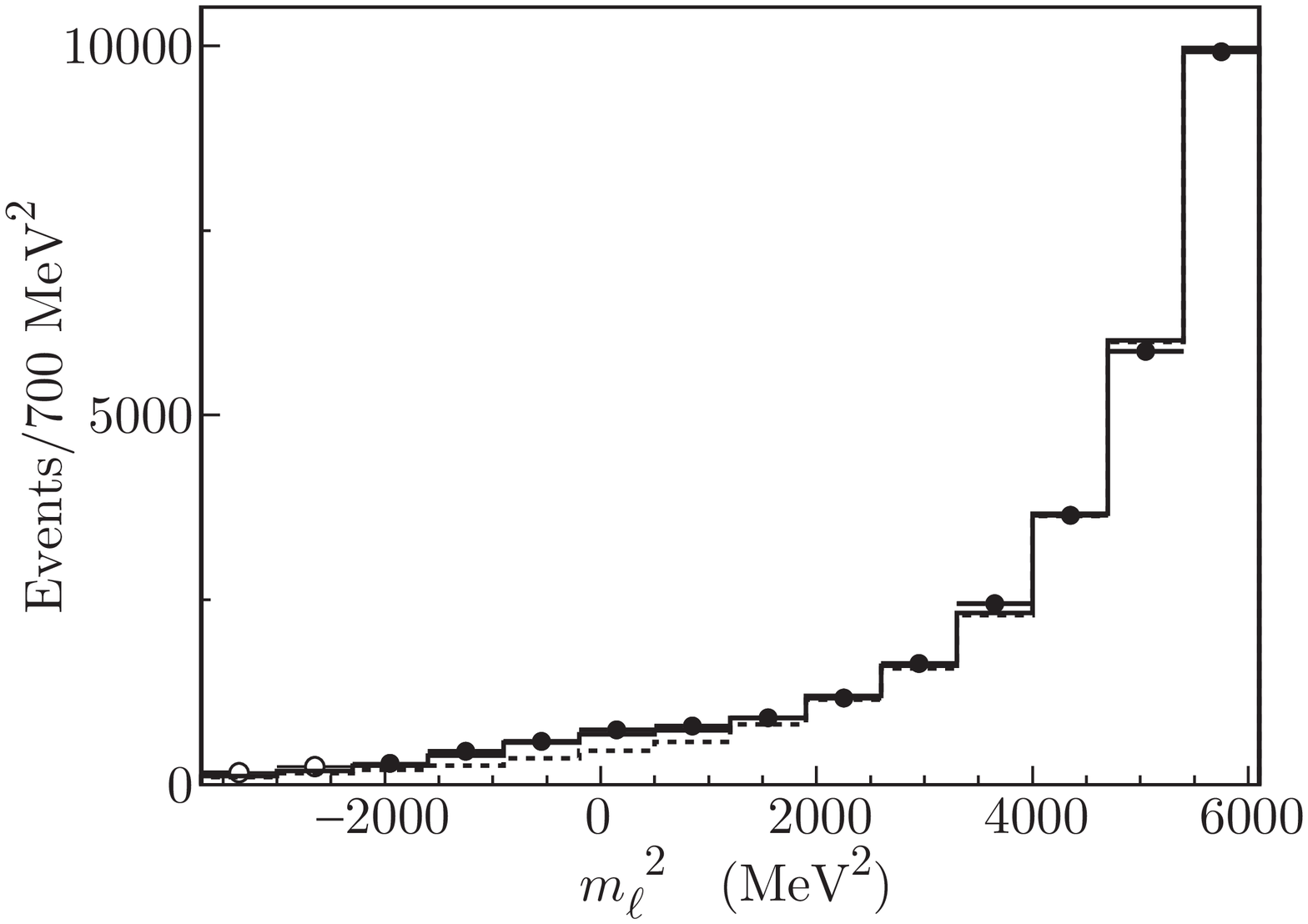}
  \caption{Fit projections onto the $m_\ell^2$ axis for $\nn>0.98$ (left) and $\nn<0.98$ (right), for data (black dots), MC fit (solid line), and  \kmd\ background (dotted line). The contribution from \ked\ events
with $E_\gamma>10$ MeV is visible in the left panel (dashed line).}
  \label{fig:fitke2}
\end{figure*}

We count 7064$\pm$102 $K^+\to e^+\nu(\gamma)$ events and 6750$\pm$101 $K^-\to e^-\bar{\nu}(\gamma)$,
89.8\% of which have  $E_\gamma<10$ MeV.
The signal-to-background correlation is $\sim 20\%$ and the
$\chi^2/\mathrm{ndf}$ is 113/112 for $K^+$ and 140/112 for $K^-$.
Fig. \ref{fig:fitke2} shows the sum of fit results for $K^+$ and
$K^-$ projected onto the $m_\ell^2$ axis in a signal
($\nn>0.98$) and a background ($\nn<0.98$) region.

The number of \kmd\ events is obtained from a fit to the $m_\ell^2$ distribution.  The fraction of
background events under the muon peak is estimated from MC to be less
than one per mil. We count $2.878\times10^8$ ($2.742\times10^8$)
$K^+\to \mu^+\nu(\gamma)$ ($K^-\to \mu^-\bar{\nu}(\gamma)$) events.  The difference between $K^+$ and
$K^-$ counts is due to $K^-$ nuclear interactions in the material traversed.

\section{Efficiency and systematic errors evaluation}
\label{sec:cs}
The ratios of \ked\ to \kmd\ efficiencies
are evaluated with MC and corrected for possible differences between data and MC, using control samples.
We evaluate data-MC corrections separately for each of the following analysis steps: 
decay point reconstruction (kink), quality cuts, cluster-charged particle association. 
For each step, the correction is defined as the ratio of data and MC efficiencies measured on the 
control sample, each folded with the proper kinematic spectrum of \ked\ (or \kmd) events.

Decay point reconstruction efficiencies are evaluated using pure samples of \kmd\ and
\ket; these are tagged by the identification of the two-body decay, \kmd\ or $K \to \pi \pi^0$ (\kpd), 
of the other kaon 
and selected with tagging and EMC information only, without using tracking.  
The corrections to MC efficiencies range between 0.90 and 0.99 depending on the decay
point position and on the decay angle. The simulation is less accurate in case of
overlap between lepton and kaon tracks, and with decays occurring close
to the inner border of the fiducial volume.

Samples of $K_L(e3)$, $K_L(\mu 3)$, and \kmd\ decays with a purity of
99.5\%, 95.4\%, and 100.0\% respectively, are used to evaluate lepton cluster
efficiencies. These samples are selected using tagging and DC information only,
without using calorimeter, see Refs. \cite{klsl}. 
The efficiency has been evaluated as a function of
the particle momentum separately for barrel and endcap. The
correction to MC efficiencies ranges between 0.98 and 1.01 depending on the momentum and
on the point of impact on the calorimeter.
The trigger efficiency has been evaluated solely from data.

\def\x{\times}
\def\dsten{$\delta(R_{10})$}  \def\dsgam{$\delta(R_{\gam})$}


The absolute values of all of the systematic uncertainties on $R_{10}$
are listed in Table \ref{tab:syske2}.
All of the sources of systematic error are discussed below.
\begin{table}[ht!]
  \begin{center}
    \begin{tabular}[c]{|ccc|}\hline
                          & &\vp \dsten$\x10^5$ \\ \hline
    \multicolumn{2}{|c}{Statistical error}                 & 0.024  \\ \hline
    \multicolumn{2}{|c}{Systematic error}                  &        \\
            Counting: &    fit                             & 0.007  \\
                      &     DE                             & 0.005  \\
            Efficiency:& kink                              & 0.014  \\
                & trigger                                  & 0.009  \\
                & $e, \mu$ cluster                         & 0.005  \\
               \multicolumn{2}{|c}{Total systematic error} & 0.019  \\ \hline
    \end{tabular}
  \end{center}
  \caption{Summary of statistical and systematic uncertainties on the measurements of $R_{10}$.}
  \label{tab:syske2}
\end{table}

To minimize possible biases on \ked\ event counting due to the
limited knowledge of the momentum resolution, we used \kmd\ data to carefully tune the MC response on the
tails of the $m_\ell^2$ distribution. This has been performed in sidebands of the \nn\ variable, to
avoid bias due to the presence of \ked\ signal.
Similarly, for the \nn\ distribution the EMC response in the MC has been tuned at the level of single cell,
using $K_{\ell 3}$ data control samples. 
Residual differences between data and MC \ked\ and \kmd\ \nn\ shapes have been corrected by using 
the same control samples.
Finally, to evaluate the systematic error associated with these procedures, 
we studied the variation of the results
with different choices of fit range, corresponding to a change of overall purity from
$\sim75\%$ to $\sim10\%$, for $\ke(\gamma)$ with $E_\gamma<10$ MeV.
A systematic uncertainty of $\sim0.3\%$ is derived by scaling the uncorrelated errors 
so that the reduced $\chi^2$ value equals unity
(see also Table \ref{tab:syske2}).

\ked\ event counting is also affected by the uncertainty on $f_\mathrm{DE}$, the fraction
of \ked\ events in the fit region which are due to DE process. 
This error has been evaluated by repeating the measurement of
$R_{10}$ with values of $f_\mathrm{DE}$ varied within its uncertainty, 
which is $\sim 4\%$ according to our measurement
of the \kedg\ differential spectrum \cite{kaon09:moulson}.
Since the $m_\ell^2$ distributions for \kedg\ with $E_\gamma<10$ MeV and with $E_\gamma>10$ MeV
overlap only partially, the associated fractional variation on $R_{10}$ is reduced:
the final error due to DE uncertainty is 0.2\% (Table \ref{tab:syske2}).

Different contributions to the systematic uncertainty on
$\epsilon_{e2}/\epsilon_{\mu2}$ are listed in Table \ref{tab:syske2}.
These errors are do\-mi\-na\-ted by the statistics of the control samples used to
correct the MC evaluations.  In addition, we studied the variation
of each correction with modified control-sample selection criteria. We
found neglible contributions in all cases but for the kink
and quality cuts corrections, for which the bias due to the control-sample selection
and the statistics contribute at the same level.

The total systematic error is $\sim0.8\%$, to be compared
with statistical accuracy at the level of $\sim 1\%$.

\section{\mbo $R_K$ and lepton-flavor violation}
\label{sec:results}

The number of $\ke(\gamma)$ events with $E_\gamma<10$ MeV,
the number of $\km(\gamma)$ events, the ratio of \ked\ to \kmd\ efficiencies
and the measurement of $R_{10}$ are given in Table \ref{tab:fitresultske2} for
$K^+$, $K^-$ and both charges combined.
$K^+$ and $K^-$ results are consistent within the statistical error. The systematic uncertainty is
common to both charges.
\begin{table*}[ht!]
  \begin{center}
    \begin{tabular}[c]{|c|c|c|c|c|}\hline
         & N($K_{e2}$) &   N($K_{\mu2}$)   &  $\epsilon_{e2}/\epsilon_{\mu2}$ & $R_{10}\times10^{5}$  \\ \hline
   $K^+$ &$6348\pm92\pm23$&$2.878\times10^8$&$0.944\pm0.003\pm0.007$&$(2.336\pm0.033\pm0.019)   $\\
   $K^-$ &$6064\pm91\pm22$&$2.742\times10^8$&$0.949\pm0.002\pm0.007$&$(2.330\pm0.035\pm0.019)   $\\
   $K^\pm$&$12412\pm129\pm45$&$5.620\times10^8$&$0.947\pm0.002\pm0.007$&$(2.333\pm0.024\pm0.019)$\\ \hline
    \end{tabular}
  \end{center}
  \caption{Number of \ked\ and \kmd\ events, efficiency ratios and results for $R_{10}$
   for $K^+$, $K^-$, and both charges combined; first error is statistical, second one is systematic.}
  \label{tab:fitresultske2}
\end{table*}

\begin{figure}[ht!]\centering
  \includegraphics[width=0.3\linewidth]{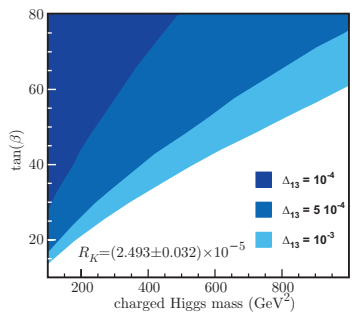}
  \caption{Excluded regions at 95\% C.L. in the plane
 $M_H$--$\tan \beta$ for $\Delta_R^{31}=10^{-4}, 5\times 10^{-3},10^{-3}$.}
  \label{fig:rkmssm}
\end{figure}

To compare the $R_{10}$ measurement with the inclusive $R_K$ prediction from SM, we take
into account the acceptance of the 10 MeV cut for IB, Eq. \ref{eq:R10new}. We obtain:
$R_K = (2.493\pm0.025_\mathrm{stat}\pm0.019_\mathrm{syst})\times 10^{-5}$,
in agreement with SM prediction of Eq. \ref{eq:rksm}. 
In the framework of MSSM with lepton-flavor violating (LFV)
couplings, $R_K$ can be used to set constraints in the space of relevant parameters, 
using the following expression \cite{masiero}:
\be
\label{eq:rkmssm}
  R_K=R_K^{\mathrm{SM}}\times\left[1+ \left(\frac{m_K^4}{m_H^4}\right)
  \left(\frac{m^2_\tau}{m^2_e}\right)
  \left|\Delta_R^{31}\right|^2\tan^6\beta\right], \ee
where $M_H$ is the charged-Higgs mass, $\Delta_R^{31}$ is the effective $e$-$\tau$
coupling constant depending on MSSM parameters, and $\tan\beta$ is the
ratio of the two Higgs superfields vacuum expectation values.
The regions excluded at 95\% C.L. in the plane $M_H$--$\tan\beta$ are shown
in Fig. \ref{fig:rkmssm} for different values of the effective LFV
coupling $\Delta_R^{31}$.

\section{Conclusions}
We have performed a comprehensive study of the process \kedg. We have measured
the ratio of \kedg\ and \kmd\ widths for photon energies smaller than 10 MeV, without photon detection
requirement.  We find:
$R_{10}=(2.333\pm0.024_\mathrm{stat}\pm0.019_\mathrm{stat})\times 10^{-5}$.
From this result we derive the inclusive ratio $R_K$ to be compared with the SM prediction:
$R_K=(2.493\pm0.025_\mathrm{stat}\pm0.019_\mathrm{syst})\times 10^{-5}$,
in excellent agreement with the SM prediction
$R_K= (2.477\pm0.001)\times 10^{-5}$.
Our result improves the accuracy with which $R_K$ is known by a factor of 5 with respect to the
present world average and allows severe constraints to be set
on new physics contributions in the MSSM with lepton flavor violating couplings.

\end{document}